\documentclass{article}

\usepackage[nonatbib, final]{neurips_2025_ml4ps}

\usepackage[utf8]{inputenc} % allow utf-8 input
\usepackage[T1]{fontenc}    % use 8-bit T1 fonts
\usepackage{hyperref}       % hyperlinks
\usepackage{url}            % simple URL typesetting
\usepackage{booktabs}       % professional-quality tables
\usepackage{amsfonts}       % blackboard math symbols
\usepackage{nicefrac}       % compact symbols for 1/2, etc.
\usepackage{microtype}      % microtypography
\usepackage{xcolor}         % colors
\usepackage{amsmath}        % math formatting
\usepackage{graphicx}
\usepackage{wrapfig}
\usepackage[backend=biber]{biblatex}
\title{Combining datasets with different ground truths using Low-Rank Adaptation to generalize image-based CNN models for photometric redshift prediction
}

\author{
  Vikram Seenivasan$^{1}$\\
  \texttt{vikrams25@ucla.edu} \\
  \And 
  Srinath Saikrishnan$^{2}$\\
  \texttt{srinathsai22@ucla.edu} \\
  \And
  Andrew Lizarraga$^{3}$\\
  \texttt{andrewlizarraga@ucla.edu} \\
  \And
  Jonathan Soriano$^{1}$\\ \texttt{jsoriano@astro.ucla.edu} \\
  \And
  Bernie Boscoe$^{4}$\\
  \texttt{boscoeb@sou.edu} \\
  \And 
  Tuan Do$^{1}$\\
  \texttt{tdo@astro.ucla.edu} 
}

\begin{document}

\maketitle

\begin{center}
$^{1}$ Department of Physics and Astronomy, UCLA, Los Angeles, CA 90095 \\
$^{2}$ Department of Computer Science, UCLA, Los Angeles, CA 90095 \\
$^{3}$ Department of Statistics and Data Science, UCLA, Los Angeles, CA 90095 \\
$^{4}$ Department of Computer Science, Southern Oregon University,  Ashland, OR, 97520 \\

\end{center}

\begin{abstract}
  % The abstract paragraph should be indented \nicefrac{1}{2}~inch (3~picas) on
  % both the left- and right-hand margins. Use 10~point type, with a vertical
  % spacing (leading) of 11~points.  The word \textbf{Abstract} must be centered,
  % bold, and in point size 12. Two line spaces precede the abstract. The abstract
  % must be limited to one paragraph.

  In this work, we demonstrate how Low-Rank Adaptation (LoRA) can be used to combine different galaxy imaging datasets to improve redshift estimation with CNN models for cosmology. LoRA is an established technique for large language models that adds adapter networks to adjust model weights and biases to efficiently fine-tune large base models without retraining. We train a base model using a photometric redshift ground truth dataset, which contains broad galaxy types but is less accurate. We then fine-tune using LoRA on a spectroscopic redshift ground truth dataset. These redshifts are more accurate but limited to bright galaxies and take orders of magnitude more time to obtain, so are less available for large surveys. Ideally, the combination of the two datasets would yield more accurate models that generalize well. The LoRA model performs better than a traditional transfer learning method, with $\sim2.5\times$ less bias and $\sim$2.2$\times$ less scatter. Retraining the model on a combined dataset yields a model that generalizes better than LoRA but at a cost of greater computation time. Our work shows that LoRA is useful for fine-tuning regression models in astrophysics by providing a middle ground between full retraining and no retraining. LoRA shows potential in allowing us to leverage existing pretrained astrophysical models, especially for data sparse tasks. 

\end{abstract}

\section{Introduction}
Astronomy is in the era of Big Data, with several current and upcoming surveys such as LSST \cite{ivezic2019b} and Euclid \cite{laureijs2011a} imaging billions of galaxies to constrain cosmological models governing the origin and evolution of the universe. Galaxy redshifts are used to determine distances to galaxies, allowing us to map the structure of the universe through cosmic time, to determine properties of dark energy and dark matter that govern the accelerated expansion of the universe. Astrophysicists are increasingly adopting machine learning techniques to efficiently make measurements, such as galaxy redshifts, from large datasets \cite{salvato2019, newman2022b}. Such measurements often require leveraging different sources of ground truth. In this paper, we explore using Low-Rank adaptation (LoRA)\cite{hu2021} and traditional transfer learning to fine-tune redshift prediction models for galaxies by integrating datasets with different sources of ground truth. 

\subsection{Related Works}

\textbf{Redshift estimation:} Traditionally, redshifts are obtained using spectroscopy. These are highly accurate but limited to bright galaxies that have strong emission lines, and are not available at the scale of large upcoming surveys as they are expensive \cite{newman2022b}. Photometric redshifts are derived from galaxy
\textit{photometry} (brightness) in different wavelength bands to sample spectral properties. These are less accurate \cite{beck2016} but enable the analysis of much larger datasets. Recently, many machine learning models have been proposed to estimate photometric redshifts \cite{collister2004,carrascokind2013, bonnett2015,beck2016,sadeh2016,jones2017,jones2024,salvato2019,schuldt2021}. For example, convolution neural networks (CNNs) produce some of the most precise redshift estimates \cite{pasquet2019,schuldt2021, dey2022,jones2024a} by using galaxy images directly. These methods usually rely on spectroscopic redshift as ground truth for training. Recent work by \cite{cabayol2023,soriano2024} explored methods of incorporating photometric redshifts from multi-band photometry as a different source of ground truth. However, these methods are trained on photometry alone, losing out on the wealth of information in images \cite{lizarraga2024}. 

\textbf{LoRA and fine-tuning:} LoRA  is an established technique for fine-tuning large language models that adds adapter networks with lower-rank weight matrices to pre-trained models. This allows efficient fine-tuning of large base models, with only a tiny fraction of the parameters being updated \cite{hu2021}. Low-Rank adaptation has been used in astrophysics to fine-tune pre-trained models from other domains for astrophysics tasks. For example, LLMs for token-based redshift estimation \cite{ramachandra2025a}, foundation models for radio astronomy \cite{riggi2025}, and diffusion models for galaxy image generation \cite{bishnoi2025}. Its utility for regression-based tasks is not fully explored in astrophysics.

\subsection{Contribution}

We explore using LoRA and transfer learning to integrate different sources of ground truth for a redshift regression CNN that takes 5-band galaxy images as input. We adapt the recently developed LoRA implementations \cite{peft} for CNNs for astrophysical data. To our knowledge, this work is the first application of LoRA for  regression-based tasks for astrophysics. Previous work in astrophysics used pre-trained models from other domains---this is the first work that uses baseline models trained on astronomical data. These are key considerations vital for large-scale surveys that will leverage previous astrophysics knowledge.

\section{Data}
\label{sec:data}
In this work we use three datasets, one for the base model, one for transfer learning, and a combination dataset of the two. The dataset used to train the base model is derived from TransferZ \cite{soriano2024} with the addition of 5-band images from the Hyper Suprime-Cam Subaru Strategic Program (HSC-SSP) Second Public Data Release \cite{aihara2019b} and additional quality cuts as specified in \cite{do2024b}. 
\begin{wrapfigure}{r}{0.4\textwidth}
 \centering \includegraphics[width=\linewidth]{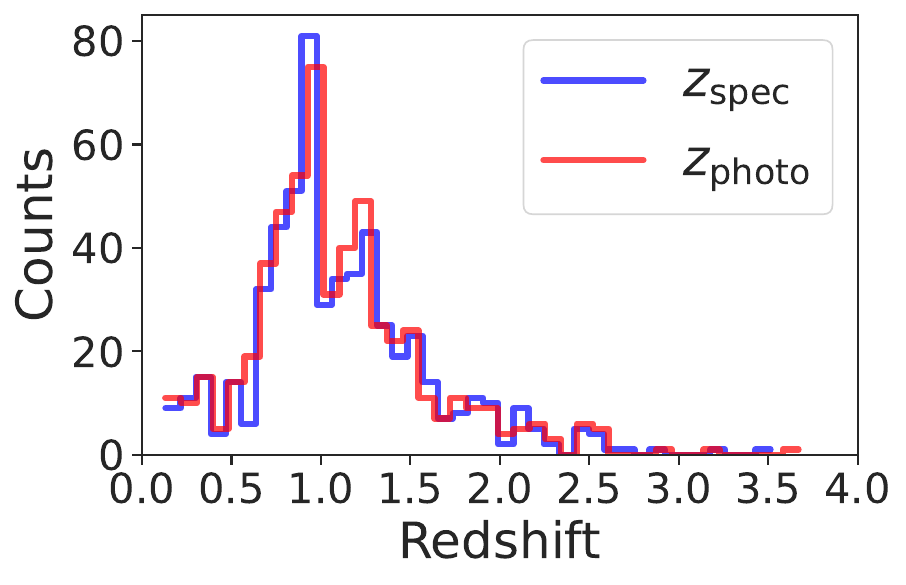}
 \caption{The redshift number distributions of the 557 common galaxies using spectroscopic ground truth from GalaxiesML (blue) and photometric ground truth from TranferZ-Images (red). }
 \label{fig:common_galaxies}
\end{wrapfigure}
This results in a dataset we call TransferZ-Images \footnote{TransferZ-Images is available on Zenodo \href{https://zenodo.org/records/16989604?preview=1&token=eyJhbGciOiJIUzUxMiJ9.eyJpZCI6ImFlOThiYmM0LTRlMGItNDgzNS1iNjlkLTZmOWM1NzMyYmI5NyIsImRhdGEiOnt9LCJyYW5kb20iOiIwNGJjNjk1MDM5NTI5YWFmYjFmMDJjYjlhMTUzMzRlNCJ9.CQMk2RAOmlEFDWiu6ln6uQPc8Q_-0GTfzw47p_4EBz9ZjZjGJI_lrgpdSF-nwfrVboTfMOdBfnHe-tw8nPI5Xg}{here.}}  consisting of 100,442 galaxies with 5-band \textit{g, r, i, z, y} images. We use the GalaxiesML\cite{do2024b} \footnote{GalaxiesML is available on Zenodo \href{https://zenodo.org/records/11117528}{here.}}dataset for fine-tuning. These datasets have been used for redshift estimation in the context of cosmology, and are ML-ready and complement each other by covering different galaxy color types and redshift ranges. 

The TransferZ-Images dataset has redshift ground truths from photometric redshift estimates while GalaxiesML has ground truth from spectroscopy. The TransferZ-Images photometric redshifts are less accurate than GalaxiesML but contain a fainter and larger variation in galaxy types. Ideally, it is a good dataset to train the base model. We aim to fine-tune on GalaxiesML to improve the base model with more accurate redshifts. In principle, the combination should produce a model that is more accurate and generalizable. We additionally create the Combo dataset by combining TransferZ-Images and GalaxiesML, resulting in a dataset containing 386,286 galaxies. For the 557 common galaxies in both datasets we use the GalaxiesML spectroscopic redshift as ground truth. Additionally, comparing the spectroscopic and photometric ground truth redshifts for the overlapping galaxies, we find no systematic bias in both redshift distributions (see Fig. \ref{fig:common_galaxies}). The data used in this study is summarized in Table \ref{data-table}. The three datasets are split into 80$\%$ training, 10$\%$ validation and 10$\%$ testing subsets.

\begin{table}[ht]
  \caption{Data Summary.}
  \label{data-table}
  \centering
  \begin{tabular}{lcccccc}
    \toprule
    % Dataset & Ground Truth & \# Sources & 90th \%ile \(z\) & Median \(z\) Unc. & 90th \%ile i-band mag & \# Filters \\
    Dataset & Ground & No. of & $z$ & Median $z$ & i-mag & No. \\
     & Truth $z$ & Sources   & 90th $\%$ile & Uncertainty 
    &   90th $\%$ile   & Filters \\
    \midrule
    TransferZ-Images & \(z_{\mathrm{phot}}\) & 100,442 & 1.8 & 0.03   & 25 & 5 \\
    GalaxiesML        & \(z_{\mathrm{spec}}\) & 286,401 & 1.2 & 0.0002 & 22 & 5 \\
    Combo             & \(z_{\mathrm{phot}}\) \& \(z_{\mathrm{spec}}\) & 386,286 & 1.4 & 0.0006 & 24 & 5 \\
    \bottomrule
  \end{tabular}
\end{table}

\section{Methodology and Metrics}

We train four models for this study to compare the effectiveness of LoRA to generalize redshift estimation models. These are (1) the base model (CNN-Base) trained on TransferZ; (2) the traditionally transfer learned model, built from fine-tuning CNN-base on GalaxiesML (CNN-TL); (3) the LoRA model, built from fine-tuning CNN-Base on GalaxiesML using LoRA (CNN-LoRA); and (4) the full retraining model (CNN-Combo) trained on Combo. These models share a ResNet18 architecture \cite{he2015}, modified to accept five images as input, with a regressor network attached to produce a single redshift value as output. We perform hyperparameter tuning on learning rate, batch size, and regressor architecture. The learning rate was the most sensitive parameter. The regressor consists of two linear layers with 512 and 256 neurons, with dropout layers (using standard dropout of 0.5) and SiLU activation between them to promote stable model training. 

\begin{wrapfigure}{r}{0.5\textwidth}
 \centering
 \includegraphics[width=\linewidth]{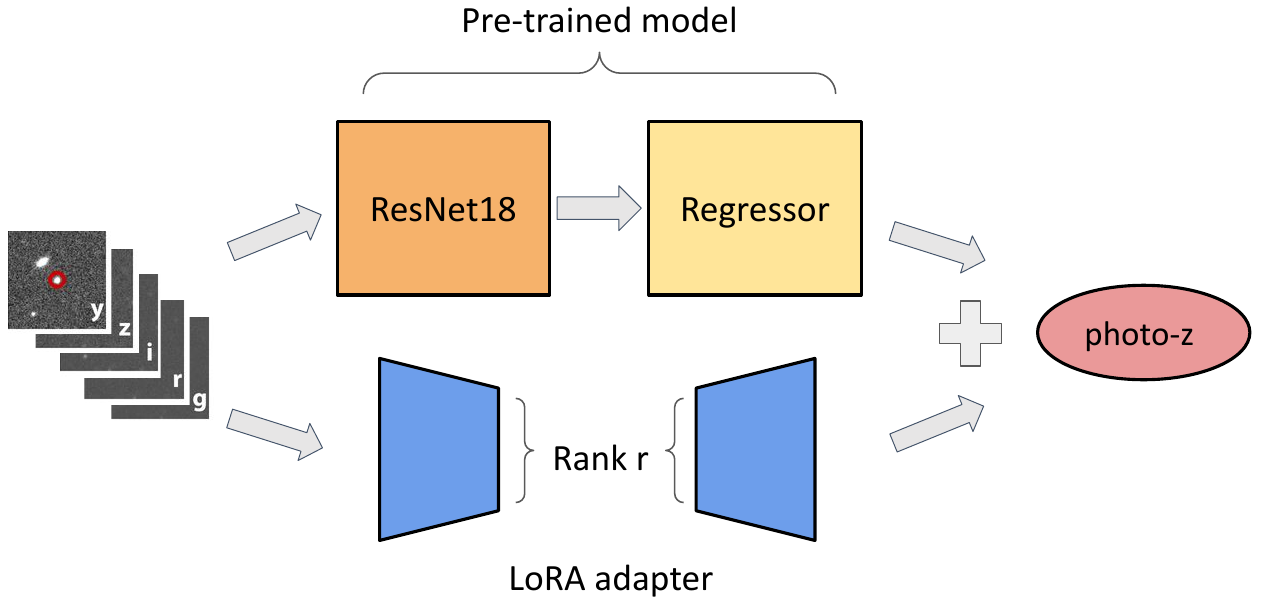}
 \caption{Visualization of LoRA implementation on a ResNet + regressor Model, based on Fig 1. in \cite{hu2021}.}
 \label{fig:lora}
\end{wrapfigure}

To train these models, we use a \texttt{ReduceLROnPlateau} scheduler with the same parameters as in \cite{he2015}, the Adam optimizer, and a custom loss function, $L(\Delta z) = 1 - \frac{1}{1 + (\Delta z/0.15)^2}$, from \cite{tanaka2018}. This loss function encapsulates photometric redshift accuracy metrics, such as outlier rate, bias and scatter which are the most important metrics for cosmology \cite{jones2024a}(see below). We use an exponential moving average (EMA) of the validation loss to determine when to stop training. The models are set to train for 500 epochs with early stopping which ends training if the model's EMA validation loss does not improve after 20 epochs. We use an AMD Ryzen Threadripper PRO 3955WX with 16-Cores and NVIDIA RTX A6000 to train the models. For CNN-Base, we use a learning rate of 1e-3, and it stops training after 56 epochs taking 26.7 minutes with a final learning rate of 1e-8. This is used as a base to train CNN-TL and CNN-LoRA. To create CNN-TL, all the model weights in CNN-Base are frozen except the input and regressor layers, and it is trained on GalaxiesML with the final learning rate of 1e-8 from the CNN-Base training. It trains for the full 500 epochs, taking 4.4 hours. To create CNN-LoRA, we use the  \texttt{peft} library (Parameter Efficient Fine-Tuning) \cite{xu2023}. We create a LoRA adapter that spans all the linear and convolutional layers in the model (Fig. \ref{fig:lora}). The adapter has a rank of 4, alpha parameter of 16 and dropout of 0.1, chosen from hyperparameter tuning. The LoRA adapter is trained with GalaxiesML, stopping in 121 epochs, taking 1.4 hours. Finally, CNN-Combo is trained with the Combo dataset and a learning rate of 1e-3 for 121 epochs without early stopping, taking 2.4 hours. We choose the same number of epochs for CNN-Combo as CNN-LoRA to quantify computational efficiency gains. All models are implemented in Pytorch \cite{paszke2019}. The code is made available through Github\href{https://github.com/astrodatalab/seenivasan_2025}{ here.}

To evaluate these models, we use the bias, scatter and catastrophic outlier rate metrics \cite{tanaka2018}. The bias metric is the median of the bias distribution defined as $b = (z_{pred} - z_{truth})/(1 + z_{truth})$ where $z_{pred}$ and $z_{truth}$ are the predicted redshift and the ground truth redshift, respectively. The scatter is defined as the median absolute deviation of the bias distribution multiplied by 1.4826. The catastrophic outlier rate is the fraction of objects satisfying $|z_{photo} - z_{truth}| > 1.0$ \cite{singal2022}. We calculate the metrics in the redshift range $0.3<z<1.5$. These metrics are important for most cosmology science applications; in particular LSST requires  bias $<|0.003|$ and scatter $ < 0.02$ for photometric redshift estimates \cite{collaboration2009}. We qualitatively evaluate model ``generalizability'' based on how well it performs across all testing datasets. 
%which is a transformation defined as $\text{EMA}_t = \alpha \cdot \text{val\_loss}_t + (1 - \alpha) \cdot \text{EMA}_{t-1}$ with initialization $ \text{EMA}_0 = \text{val\_loss}_0$. We choose $\alpha = 2/(N+1) = 2/11$, which corresponds to a weighted average of the validation loss over approximately the last $N= 10$ epochs.

\section{Results and Discussion}
\begin{figure*}[hbt!]    \includegraphics[width=1\textwidth]{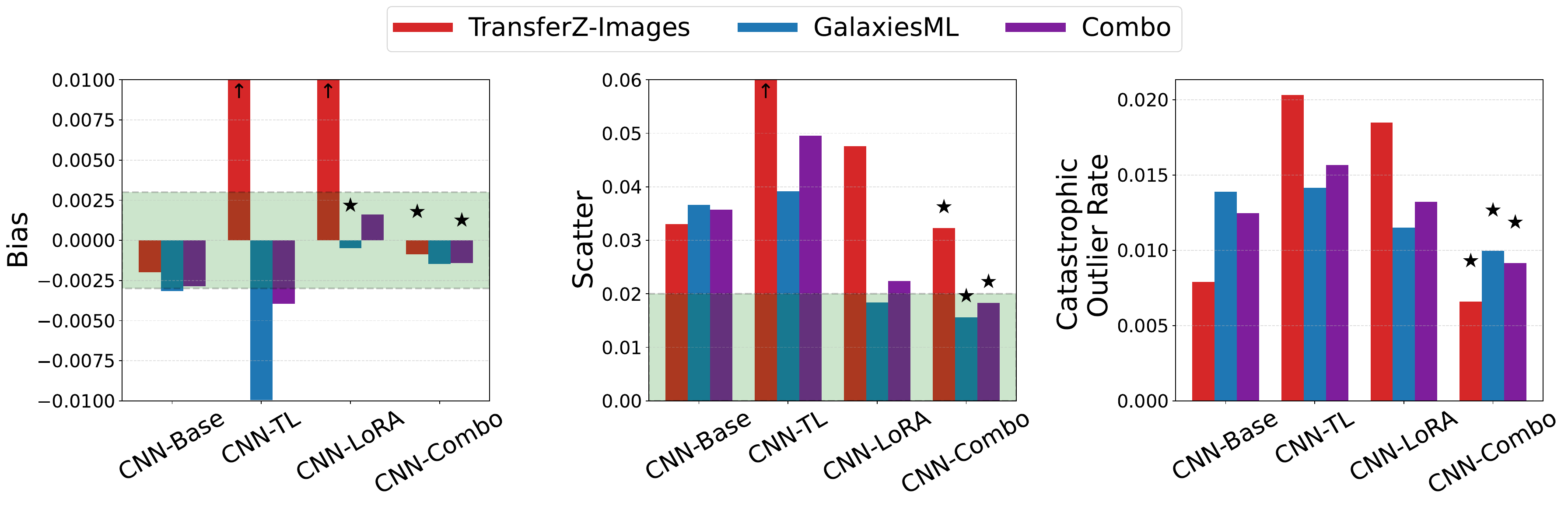}
    \caption{Bias (left), scatter (middle), and catastrophic outlier rate performance metrics (right) for all four models. The metrics are evaluated on all three datasets (TransferZ-Images (red), GalaxiesML (blue), Combo (purple)). The green band for Bias and Scatter are the LSST Science Requirements. The black stars represent the best (lowest) metric for each dataset. Upward black arrows indicates the bar overflows the plot. Low-Rank adaptation (CNN-LoRA) performs better than traditional transfer learning (CNN-TL), but not as well as retraining on the entire dataset (CNN-Combo). 
    }   \label{fig:metric_row}
\end{figure*}

We find that Low-Rank adaptation (CNN-LoRA) results in more generalizable models than traditional transfer learning (CNN-TL), but does not perform as well as retraining on combined ground truths (CNN-Combo). We evaluate the performance of the four models by testing them on the test sets of 10,027 galaxies from TransferZ-Images, 40,914 galaxies from GalaxiesML and 50,871 galaxies from Combo (See Appendix \ref{appendix_scatter_plots} for scatter plots of predictions vs. ground truth). The model performance metrics are displayed in Fig. \ref{fig:metric_row} (see Table \ref{model-performance-table} for values). CNN-LoRA has  $\sim$2.5$\times$ less bias and $\sim$2.2$\times$ less scatter on Combo compared to CNN-TL. Compared to CNN-Combo, CNN-LoRA has similar bias ($\sim$1.1$\times$ higher) and scatter ($\sim$1.2$\times$ higher) on the Combo testing set, though slightly higher. When looking at the GalaxiesML fine-tuning dataset in particular, CNN-LoRA has $\sim$3$\times$ lower bias than CNN-Combo, and $\sim$1.2$\times$ higher scatter. However, CNN-LoRA's performance on the baseline data, TransferZ-Images, is worse, with $\sim$20$\times$ higher bias and $\sim$1.5$\times$ higher scatter. This indicates that the CNN-LoRA model forgot information it learned in the baseline training on TransferZ-Images. The choice of method of mixing ground truths has a trade-off between performance requirements and computational resources available. Thus, LoRA may be a good choice if retraining the model with combined data is too computationally intensive and diminished performance on the base dataset does not detract from science goals. 

\begin{figure}[ht]    
\includegraphics[width=\linewidth]{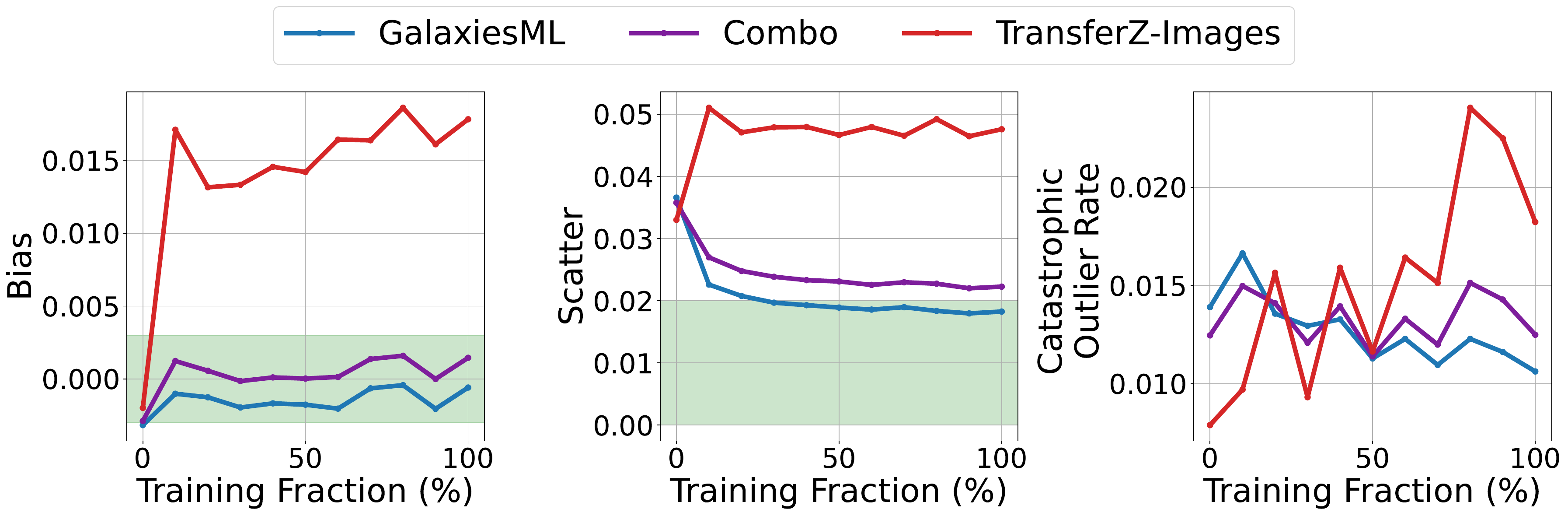}
    \caption{Bias (left), scatter (middle) and catastrophic outlier rate (right) as a function of fraction of GalaxiesML used to fine-tune using CNN-LoRA. The green band for Bias and Scatter are the LSST Science Requirements. Using just 10$\%$ of the data, the model's scatter on GalaxiesML improves and on TransferZ reduces drastically. The performance of CNN-LoRA plateaus when 40-50$\%$ of  GalaxiesML is used for training; only a fraction of the dataset is needed for the best fine-tuning.} \label{fig:metrics_vs_fraction_TZtoGM}
\end{figure}

We perform a further test of CNN-LoRA by training the model with different fractions of GalaxiesML. Training with just 10$\%$ of GalaxiesML, the model's scatter on GalaxiesML improves significantly and continues improving until about a 40-50 $\%$ data fraction (Fig. \ref{fig:metrics_vs_fraction_TZtoGM}). Beyond this, there is no significant improvement in model performance. Further, even with a 100$\%$ data fraction, CNN-LoRA trains in $\sim$1.5$\times$ less time than CNN-Combo when both models are trained to the same number of epochs. Training time is even lower with smaller data fractions. Thus, LoRA offers a way to fine-tune models with smaller fractions of data and still achieve maximum model performance on the fine-tuning dataset, all while saving computational resources, although it still performs worse than retraining on combined ground truths.

When looking at transfer learning in particular, previous work suggests that neurons are co-adapted between model layers and later models layers can be highly specialized to baseline data \cite{yosinski2014}. CNN-LoRA is able to adjust weights and biases across the entire model with its learned adapters, resulting in its better performance compared to CNN-TL which only trains on the input and regressor layers. Thus, LoRA offers a standardized way to do transfer learning on models without the need for extensive optimization of layers to unfreeze in traditional transfer learning. 

\begin{table}[htb]
 \caption{Model Metrics Summary}
 \label{model-performance-table}
 \centering
 \begin{tabular}{llrrr}
   \toprule
   \multicolumn{2}{c}{} & \multicolumn{1}{c}{Bias } & \multicolumn{1}{c}{Scatter} & \multicolumn{1}{c}{Catastrophic } \\
   \multicolumn{4}{c}{} & \multicolumn{1}{c}{Outlier Rate}  \\
   \midrule

   CNN-Base & TransferZ-Images  & -0.00198 & 0.0330 & 0.00789 \\
           & GalaxiesML & -0.00316 & 0.0366 & 0.0139 
        \\
           & Combo &-0.00287& 0.0358 & 0.0125
          
        \\
   CNN-TL  & TransferZ-Images   & 0.0502 & 0.121 &  0.0203 
   \\
           & GalaxiesML & -0.00995 & 0.0391 & 0.0141
        \\
           & Combo & -0.00396 & 0.0496 & 0.0157
           \\

   CNN-LoRA & TransferZ-Images &  0.0176 & 0.0476 &  0.0185 \\
           & GalaxiesML & -0.000481 &  0.0184 & 0.0115
        \\
           & Combo & 0.0016 & 0.0223 & 0.0132
          
        \\

   CNN-Combo & TransferZ-Images &  -0.000868 & 0.0323 &  0.00660 \\
           & GalaxiesML & -0.00146 & 0.0156 & 0.00996
        \\
           & Combo & -0.00141 & 0.0183 & 0.00915
          
        \\
    \midrule
    LSST Req. & & < |0.03| & < 0.02 & 
    \\
   \bottomrule
\end{tabular}
\end{table}

Future work tuning LoRA parameters can further optimize its performance, and potentially rival CNN-Combo in performance with lower computational costs. An ablation study freezing and unfreezing different layers to create different CNN-TL models can also further characterize traditional transfer learning performance and compare against LoRA fine-tuning. Further work can also explore training CNN-Combo using early stopping and with different data fractions to fully quantify its performance and more robustly compare to LoRA fine-tuning. Ultimately, LoRA and other fine-tuning methods will be crucial as we ramp up to current and upcoming cosmological surveys. LoRA demonstrates potential in leveraging the wide array of existing pre-trained astrophysical models for data sparse tasks, especially when retraining models is too computationally expensive, and can adapt existing models to multiple new datasets, representing a developing avenue to do new precision cosmology.

\clearpage

\begin{ack}
%Use unnumbered first level headings for the acknowledgments. All acknowledgments go at the end of the paper before the list of references. Moreover, you are required to declare funding (financial activities supporting the submitted work) and competing interests (related financial activities outside the submitted work).
%More information about this disclosure can be found at: \url{https://neurips.cc/Conferences/2025/PaperInformation/FundingDisclosure}.

%Do {\bf not} include this section in the anonymized submission, only in the final paper. You can use the \texttt{ack} environment provided in the style file to automatically hide this section in the anonymized submission.
Partial support for this work was provided by the Alfred P. Sloan Foundation and the NSF DGE2034835. This work used Jetstream2 at Indiana University through allocation \# PHY230092 from the Advanced Cyberinfrastructure Coordination Ecosystem: Services $\&$ Support (ACCESS) program, which is supported by U.S. National Science Foundation grants \#2138259, \#2138286, \#2138307, \#2137603, and \#2138296. 

The Hyper Suprime-Cam (HSC) collaboration includes the astronomical communities of Japan and Taiwan, and Princeton University. The HSC instrumentation and software were developed by the National Astronomical Observatory of Japan (NAOJ), the Kavli Institute for the Physics and Mathematics of the Universe (Kavli IPMU), the University of Tokyo, the High Energy Accelerator Research Organization (KEK), the Academia Sinica Institute for Astronomy and Astrophysics in Taiwan (ASIAA), and Princeton University. Funding was contributed by the FIRST program from the Japanese Cabinet Office, the Ministry of Education, Culture, Sports, Science and Technology (MEXT), the Japan Society for the Promotion of Science (JSPS), Japan Science and Technology Agency (JST), the Toray Science Foundation, NAOJ, Kavli IPMU, KEK, ASIAA, and Princeton University. 

This paper makes use of software developed for the Large Synoptic Survey Telescope. We thank the LSST Project for making their code available as free software at  http://dm.lsst.org

This paper is based [in part] on data collected at the Subaru Telescope and retrieved from the HSC data archive system, which is operated by the Subaru Telescope and Astronomy Data Center (ADC) at National Astronomical Observatory of Japan. Data analysis was in part carried out with the cooperation of Center for Computational Astrophysics (CfCA), National Astronomical Observatory of Japan. The Subaru Telescope is honored and grateful for the opportunity of observing the Universe from Maunakea, which has the cultural, historical and natural significance in Hawaii. 

This work is based on observations collected at the European Southern Observatory under
ESO programme ID 179.A-2005 and on data products produced by CALET and the Cambridge Astronomy Survey Unit on behalf of the UltraVISTA consortium.
\end{ack}

\printbibliography
% \bibliographystyle{abbrvnat}
% \bibliography{cnn_transfer_NeurIPS_ML4PS_2025}

\clearpage

\appendix

\section{Redshift Prediction Plots} \label{appendix_scatter_plots}
\begin{figure*}[hbt!]    
\centering
\includegraphics[width=1\textwidth]{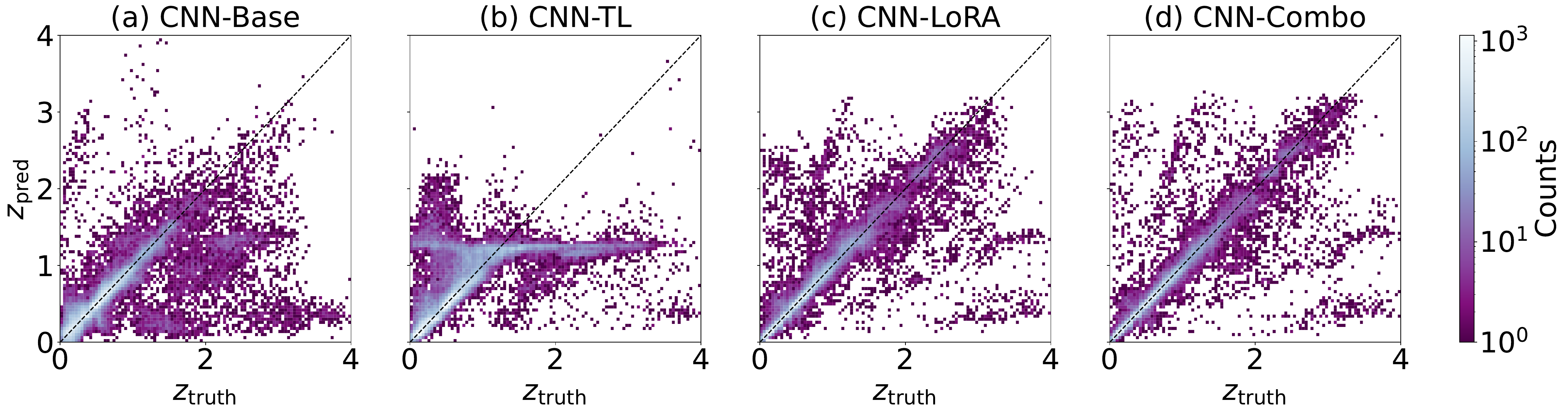}
    \caption{The true redshift vs predicted redshift for four models (a) CNN-Base, (b) CNN-TL, (c) CNN-LoRA and (d) CNN-Combo, evaluated on the Combo dataset. The color bar describes the log norm of the number of galaxies in each 2D histogram bin. The identity line (1-1 line) represents perfect agreement between ground truth and predictions. Low-Rank adaptation (CNN-LoRA) performs better than traditional transfer learning (CNN-TL), but not as well as retraining on the entire dataset (CNN-Combo).} 
    %CHANGE x axis to z_ground truth
    % label each a, b, c, d
    %change colormap to a consistent color
    \label{fig:lora_progression_combo_row}
\end{figure*}

In this section, we present the true redshift vs predicted redshift plots for the four models, CNN-Base, CNN-TL, CNN-LoRA and CNN-Combo, evaluated on the Combo dataset (Fig. \ref{fig:lora_progression_combo_row}). CNN-LoRA has improved predictions compared to CNN-Base, with lower dispersion along the 1-1 line and better predictions above $z\sim2$. CNN-LoRA and CNN-Combo display similar structures in their plots, but CNN-LoRA is more dispersed than CNN-Combo along the 1-1 line.  CNN-TL displays systematic bias, predicting redshift around $\sim$1.2--1.4 for a large number of galaxies, indicating model failure to retain knowledge or learn during transfer learning. This could be interpreted as due to a lack of galaxies above $z \sim 1$ in the fine-tuning dataset, GalaxiesML, or that further tuning of the CNN-TL architecture is needed. Furthermore, all the models have certain similar structures off the 1-1 line. Further study characterizing the sources making up those structures can reveal possible systematic bias in the data. 

\section{Fine-tuning in the `reverse'
direction} \label{appendix_reverse}

\begin{figure*}
\includegraphics[width=1\textwidth]{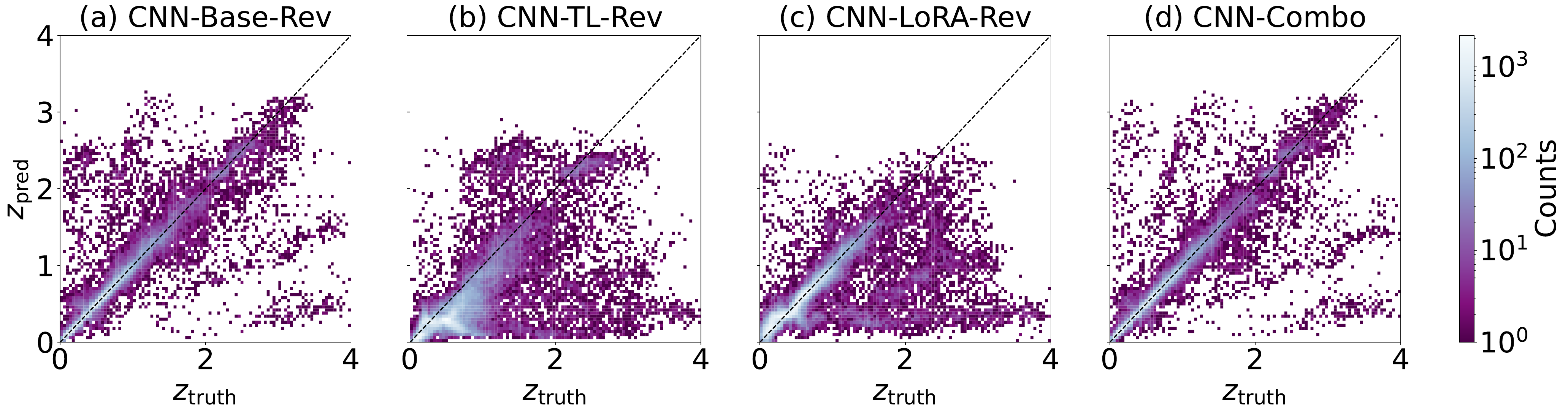}
    \caption{The true redshift vs predicted redshift for the four models (a) CNN-Base-Rev, (b) CNN-TL-Rev, (c) CNN-LoRA-Rev and (d) CNN-Combo, evaluated on the Combo dataset. The color bar describes the log norm of the number of galaxies in each 2D histogram bin. The 1-1 line represents perfect agreement between ground truth and predictions. Low-Rank adaptation (CNN-LoRA-Rev) performs better than traditional transfer learning (CNN-TL-Rev), but not as well as retraining on the entire dataset (CNN-Combo), as in the forward direction. } 
    %CHANGE x axis to z_ground truth
    % label each a, b, c, d
    %change colormap to a consistent color
    \label{fig:lora_progression_combo_row_reverse}
\end{figure*}

\begin{figure*}[hbt!]    \includegraphics[width=1\textwidth]{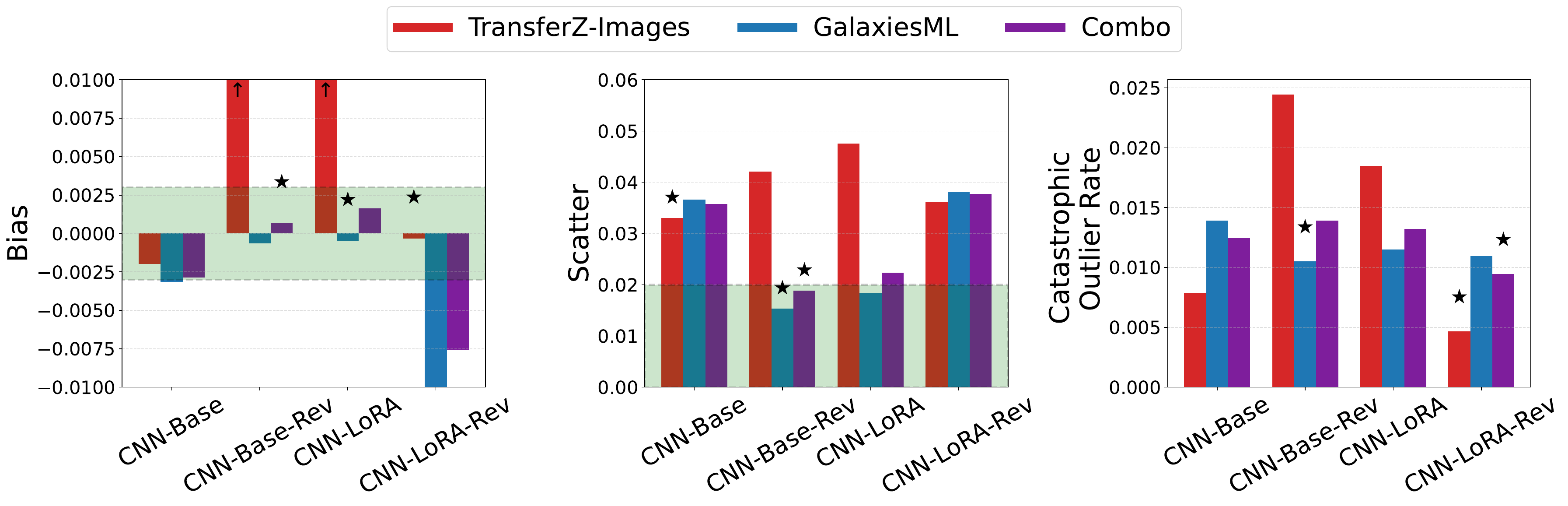}
    \caption{Bias (left), scatter (middle), and catastrophic outlier rate performance metrics (right) CNN-Base, CNN-Base-Rev, CNN-LoRA and CNN-LoRA-Rev. The metrics are evaluated on all three datasets (TransferZ-Images (red), GalaxiesML (blue), Combo (purple)). The green band for Bias and Scatter are the LSST Science Requirements. The black stars represent the best (lowest) metric for each dataset. Upward black arrows indicates the bar overflows the plot. CNN-LoRA-Rev performs worse than CNN-LoRA across most datasets. CNN-Base-Rev performs better than CNN-LoRA on scatter across all datasets, as well as bias on GalaxiesML and Combo.  
    }   \label{fig:metric_row_reverse_comparison}

\end{figure*}

To fully explore the model fine-tuning methodology, we also explored training on GalaxiesML first (CNN-Base-Rev), followed by fine-tuning with TransferZ-Images using LoRA or traditional transfer learning. CNN-Base-Rev was trained with GalaxiesML using the same architecture and hyperparameters as CNN-Base. The model  stopped training after 147 epochs taking 1.3 hours. This is longer than the CNN-Base training as GalaxiesML is a larger dataset than TransferZ-Images. We then created CNN-LoRA-Rev by fine-tuning CNN-Base-Rev using TransferZ-Images and the same LoRA configuration as CNN-LoRA. This model stopped training after 56 epochs and took 25.3 minutes. Finally, we created CNN-TL-Rev by fine-tuning CNN-Base-Rev with TransferZ-Images using the same frozen layers as for CNN-TL, which stops training after 27 epochs taking 14.7 minutes. 

Scatter plots showing the performance of four models, CNN-Base-Rev, CNN-TL-Rev, CNN-LoRA-Rev and CNN-Combo, evaluated on the Combo dataset are shown in Fig. \ref{fig:lora_progression_combo_row_reverse}. Transfer learning in the reverse direction leads to negative transfer, with CNN-TL-Rev and CNN-LoRA-Rev having more dispersion in their prediction plots and more outliers than CNN-Base-Rev. In contrast to CNN-TL discussed in Appendix \ref{appendix_scatter_plots}, CNN-TL-Rev does not fixate on predicting redshifts in the range $\sim$ 1.2--1.4, and instead under-predicts most sources with large dispersion. 

Fig. \ref{fig:metric_row_reverse_comparison} shows the bias, scatter, and catastrophic outlier rate of four models, CNN-Base, CNN-Base-Rev, CNN-LoRA and CNN-LoRA-Rev, evaluated on all test sets. Table \ref{model-performance-table-rev} contains these metrics values, and additionally, the CNN-TL, CNN-TL-Rev and CNN-Combo metrics. CNN-LoRA-Rev performs worse than CNN-LoRA on TransferZ-Images and Combo. In particular on the Combo testing set, CNN-LoRA-Rev has $\sim$4.7$\times$ higher bias and $\sim$1.7$\times$ higher scatter than CNN-LoRA. Similarly, CNN-TL-Rev performs worse than CNN-TL. Thus, the direction of integrating ground truths in these models is important, i.e. photometric redshift ground truth to spectroscopic redshift ground truth, or vice-versa. The poor transfer learning in the reverse direction can be attributed to the photometric redshift ground truths in TransferZ-Images being less precise.

Furthermore, the CNN trained on GalaxiesML alone (CNN-Base-Rev) performs better than any of the transfer learning models. For example, CNN-Base-Rev performs better than CNN-LoRA, the best of our transfer learning models, on scatter across all datasets. In particular, on the Combo testing set, CNN-LoRA has $\sim1.2\times$ higher scatter and $\sim2.4\times$ higher bias. CNN-Base-Rev in fact performs similarly to CNN-Combo on scatter on the Combo testing set (0.188 vs 0.183 in value). Thus, training a model only on spectroscopic redshift ground truth can lead to models that generalize to broader galaxy types. 

% We can additionally speculate that the higher uncertainty in photometric redshift ground truths from TransferZ-Images hinders generalizability even though the types of galaxies in the dataset are more diverse.

\begin{table}[htb]
 \caption{Model Metrics Summary Comparing Both Directions of Fine-tuning}
 \label{model-performance-table-rev}
 \centering
 \begin{tabular}{llrrr}
   \toprule
   \multicolumn{2}{c}{} & \multicolumn{1}{c}{Bias } & \multicolumn{1}{c}{Scatter} & \multicolumn{1}{c}{Catastrophic } \\
   \multicolumn{4}{c}{} & \multicolumn{1}{c}{Outlier Rate}  \\
   \midrule
   CNN-Base & TransferZ-Images  & -0.00198 & 0.0330 & 0.00789 \\
           & GalaxiesML & -0.00316 & 0.0366 & 0.0139 
        \\
           & Combo &-0.00287& 0.0358 & 0.0125
          
        \\
   CNN-Base-Rev & TransferZ-Images  & 0.0114 & 0.0421 & 0.0244 \\
           & GalaxiesML & -0.000647 &  0.0153 &  0.0105
        \\
           & Combo & 0.000672 & 0.0188 & 0.0139
          
        \\
   CNN-LoRA & TransferZ-Images &  0.0176 & 0.0476 &  0.0185 \\
           & GalaxiesML & -0.000481 &  0.0184 & 0.0115
        \\
           & Combo & 0.0016 & 0.0223 & 0.0132
        \\
   CNN-LoRA-Rev & TransferZ-Images &  -0.000322 & 0.0362 & 0.00466 \\
           & GalaxiesML & -0.0101 &  0.0381 & 0.0110
        \\
           & Combo &  -0.00759 & 0.0378 & 0.00944
          
        \\
   CNN-TL  & TransferZ-Images   & 0.0502 & 0.121 &  0.0203 
   \\
           & GalaxiesML & -0.00995 & 0.0391 & 0.0141
        \\
           & Combo & -0.00396 & 0.0496 & 0.0157
           \\
   CNN-TL-Rev & TransferZ-Images &  -0.0638 & 0.101 & 0.0259 \\
           & GalaxiesML & -0.147 &  0.0844 & 0.0149
        \\
           & Combo & -0.134 & 0.0973 & 0.0175
        \\
   CNN-Combo & TransferZ-Images &  -0.000868 & 0.0323 &  0.00660 \\
           & GalaxiesML & -0.00146 & 0.0156 & 0.00996
        \\
           & Combo & -0.00141 & 0.0183 & 0.00915
          
        \\
    \midrule
    LSST Req. & & < |0.03| & < 0.02 & 
    \\
   \bottomrule
\end{tabular}
\end{table}

\end{document}